\documentclass[prl,twocolumn]{revtex4}

\usepackage{graphics}
\usepackage{epsfig}

\begin{document}

\title{Effect of interface states on spin-dependent tunneling in Fe/MgO/Fe tunnel junctions}
\author{K. D. Belashchenko}
\author{J. Velev}
\author{E. Y. Tsymbal}
\affiliation{Department of Physics and Astronomy and Center for
Materials Research and Analysis, University of Nebraska,
Lincoln, Nebraska 68588, USA}

\begin{abstract}
The electronic structure and spin-dependent tunneling in
epitaxial Fe/MgO/Fe(001) tunnel junctions are studied using
first-principles calculations. For small MgO barrier thickness
the minority-spin resonant bands at the two interfaces make a
significant contribution to the tunneling conductance for the
antiparallel magnetization, whereas these bands are, in
practice, mismatched by disorder and/or small applied bias for
the parallel magnetization. This explains the experimentally
observed decrease in tunneling magnetoresistance (TMR) for thin
MgO barriers. We predict that a monolayer of Ag epitaxially
deposited at the interface between Fe and MgO suppresses
tunneling through the interface band and may thus be used to
enhance the TMR for thin barriers.
\end{abstract}

\pacs{72.25.Mk 73.40.Gk 73.40.Rw 73.23.-b} \maketitle

Magnetic tunnel junctions (MTJs) are miniature devices which
consist of two ferromagnetic electrodes separated by an
insulating barrier. These junctions are made in such a way that
their magnetization  may be switched between parallel and
antiparallel states under the influence of external magnetic
field. This switching is accompanied by an abrupt change of the
electric conductance of the MTJ \cite{MT}. MTJs aroused much
attention due to their potential application in magnetic
random-access memories and magnetic field sensors. In practical
terms, the figure of merit is the tunneling magnetoresistance
(TMR), which is defined by
$\mathrm{TMR}=(G_\mathrm{P}-G_{\mathrm{AP}})/G_{\mathrm{AP}}$,
where $G_\mathrm{P}$ and $G_{\mathrm{AP}}$ are the conductances
measured when the electrodes are magnetized parallel or
antiparallel to each other. Recent reviews of spin-dependent
tunneling in MTJs may be found in Refs.
\onlinecite{TML,Butler-rev}.

Since the first observation of reproducible TMR \cite{Moodera},
the majority of measurements were performed for amorphous or
polycrystalline barriers, most commonly Al$_2$O$_3$. The highest
TMR values achieved for Al$_2$O$_3$ barriers were about 70\% at
room temperature \cite{Wang}. Meanwhile, theoretical
calculations based on layer KKR \cite{Butler} and tight-binding
methods \cite{Mathon} predicted that much larger TMR values may
be obtained for coherent tunneling in epitaxial Fe/MgO/Fe(001)
junctions due to strong spin filtering. The latter is enforced
by the wave-function symmetry and its relation to the complex
band structure of the barrier \cite{Mavro}. Very large TMR
values exceeding 200\% were indeed measured for such junctions
by Parkin \emph{et al.} \cite{Parkin} and Yuasa \emph{et al.}
\cite{Yuasa}. Recently, a more accurate calculation
\cite{Blugel} based on the FLAPW method confirmed the
conclusions of Refs. \onlinecite{Butler,Mathon}.

For device applications of MTJs it is critical to make the
tunneling barrier as thin as possible in order to match the
resistance of MTJs to other electronic components. Measurements
for epitaxial Fe/MgO/Fe junctions show, however, that TMR
decreases precipitously for barrier thickness below 2~nm
\cite{Yuasa}. A detailed characterization of the MgO structure
grown on Fe(001) single crystals demonstrates a pseudomorphic
growth of MgO up to 6 monolayers (ML) ($\approx $ 1.2~nm), with
misfit dislocations being formed for thicker films \cite{Wulf}.
The two latter experimental observations suggest that in the
range of MgO thickness at which one might expect a ballistic
tunneling mechanism for conduction with no contribution from
defect scattering, TMR drops down with decreasing the barrier
thickness. The origin of this behavior is unknown. Also, these
experimental facts are in disagreement with large values of TMR
calculated for thin MgO barriers \cite{Butler,Mathon}.

In this paper we demonstrate that the reduction of TMR in
epitaxial Fe/MgO/Fe(001) junctions at small barrier thickness is
controlled by the minority-spin interface band. The presence of
this band was experimentally proven by Tiusan \emph{et al.}
\cite{Tiusan}. We show that the transmission through this
resonant channel is enhanced dramatically at small barrier
thickness making a large contribution to the conductance in the
antiparallel configuration and to the minority-spin conductance
in the parallel configuration. The latter is, however, so
sensitive to the mismatch in the potential at the two interfaces
that it is, in practice, destroyed by disorder and/or applied
bias. This explains the sizable decrease in TMR for thin MgO
barriers which is observed experimentally \cite{Yuasa}. We
predict that a monolayer of Ag epitaxially deposited at the
interface between Fe and MgO suppresses tunneling through this
interface band and may thus be used to enhance TMR for thin
barriers. This provides a new way to make MTJs with a low
resistance and high TMR that are required for device
applications. In addition, Ag interlayers protect the
ferromagnetic electrodes from oxidation which is detrimental to
TMR \cite{oxidation}.

We calculate the electronic structure and tunneling conductance
of Fe/MgO/Fe(001) MTJs with or without Ag interlayers using a
tight-binding linear muffin-tin orbital method (TB-LMTO) in the
atomic sphere approximation (ASA) \cite{Andersen} and the local
density approximation (LDA) for the exchange-correlation energy.
We use a full-potential LMTO (FP-LMTO) method \cite{FPLMTO} to
check the correctness of the ASA in describing the band
structure of the MTJ. The principal-layer Green's function
technique is applied to calculate the conductance
\cite{Turek-book}. The atomic structure of the Fe/MgO/Fe
junctions is taken from Ref.~\onlinecite{Butler}. To represent
the electronic structure within the ASA we use atomic spheres as
described in Ref. \onlinecite{atomic spheres}. The quality of
this choice of the spheres is tested against our FP-LMTO
calculations \cite{CCOR-note}. In general, we find very good
agreement between the ASA and FP results. In particular, the
band offset between Fe and MgO is reproduced very well. The
Fermi level lies approximately 3.4 eV above the MgO valence band
maximum.

\begin{figure}
\epsfig{file=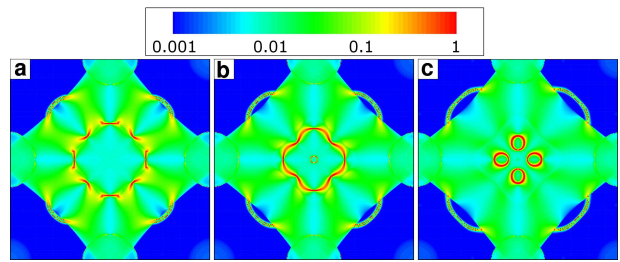,width=0.45\textwidth,clip}
\caption{Normalized minority-spin
$\mathbf{k}_\parallel$-resolved DOS at the interfacial Fe layer
in Fe/MgO/Fe(001) MTJ for three values of energy: (a) 0.02~eV
below $E_F$, (b) at $E_F$, and (c) 0.02~eV above $E_F$. The
scale is logarithmic.} \label{DOSmin}
\end{figure}

The presence of the interface band can be visualized using the
local density of states (DOS) resolved by transverse wave vector
$\mathbf{k}_\parallel$. Fig.~\ref{DOSmin} shows the
minority-spin $\mathbf{k}_\parallel$-resolved DOS at and around
the Fermi energy, $E_F$, for the interface Fe layer in a
Fe/MgO/Fe(001) junction. The band of states which is clearly
seen in red in this figure is the interface band, absent in bulk
Fe. This interface band can also be seen in the energy-resolved
DOS as the narrow peak near the Fermi level for minority-spin
electrons (see, e.g., Fig.~1a in Ref. \onlinecite{Tiusan} and
Fig.~3b in Ref.~\onlinecite{Butler}).

It is known that properties of interface (surface) states depend
on whether they are coupled to the bulk states or not
\cite{Liebsch}. In Fig.~\ref{DOSmin} the interface states located
about one quarter of the Brillouin zone width from the
$\bar\Gamma$ point are interface resonances: They lie within the
continuum of bulk Bloch states and therefore have a finite
linewidth. On the other hand, the states forming two parallel
curves in the corners of the Brillouin zone (inside the dark blue
regions) are pure interface states. In the dark blue regions the
DOS is zero in the bulk, and the interface states have zero
linewidth. To resolve these states, we added an imaginary part of
10$^{-5}$ Ry to the energy. The two parallel bands correspond to
bonding and antibonding combinations of the interface states
localized at the two sides of the barrier \cite{Wunnicke}. Near
the points where these bands enter the bulk continuum and become
resonances one can see strong peaks in the interface DOS, similar
to those predicted within a simple tight-binding model
\cite{MMM2004}.

In a single-particle approximation the interface states projected
into bulk band gaps do not contribute to the tunneling
conductance. A possible way to include the contribution of such
states was suggested by Ishida \emph{et al.} \cite{Ishida}. In our
case, however, the interface resonances lie much closer to the
$\bar\Gamma$ point compared to the pure interface states.
Therefore, the resonances dominate the conductance, and the use of
the single-particle approximation does not lead to appreciable
errors.

A notable feature of the interface band at the Fe/MgO interface is
its weak dispersion. This causes a significant change in the
location of this band within the first Brillouin zone when energy
is shifted by a tiny amount of 0.02~eV, as is seen in
Fig.~\ref{DOSmin}. This feature makes any calculation of the
interface states in Fe/MgO/Fe unreliable in terms of their Fermi
level intercepts: The LDA itself does not provide 0.01~eV
accuracy. It is very likely that this particular feature of the
interface states is the reason why earlier calculations based on
different methods \cite{Butler,Mathon,Blugel} result in very
dissimilar shapes of the minority-spin conductance plotted as a
function of $\mathbf{k}_\parallel$.

\begin{figure}
\epsfig{file=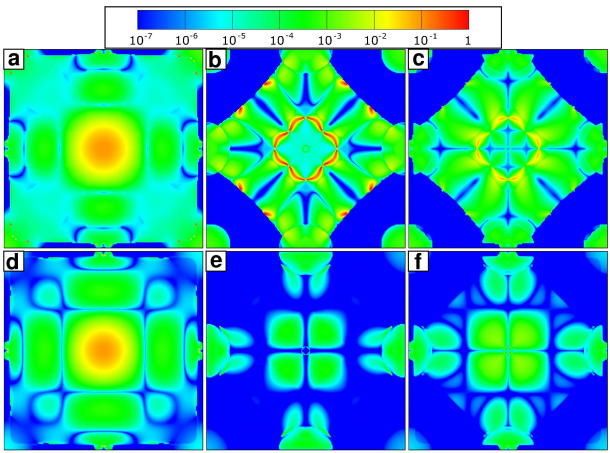,angle=270,width=0.45\textwidth,clip}
\caption{Transmission probability as a function of
$\mathbf{k}_\parallel$. (a)-(c): Fe/MgO/Fe junction with 4 MgO
MLs; (d)-(f): Same junction with Ag interlayers.  (a), (d):
Majority spins; (b), (e): Minority spins; (c), (e): Each spin
channel in the antiparallel configuration.} \label{transm}
\end{figure}

Panels (a)-(c) in Fig. \ref{transm} show the spin-resolved
transmission for the Fe/MgO/Fe junction with 4 MLs of MgO for
parallel and antiparallel magnetization. As is clearly seen from
panel (b), the resonant interface band enhances the transmission
in the minority spin channel. This enhancement is most pronounced
for small barrier thickness, because the interface band lies away
from the $\bar\Gamma$ point, and therefore the resonant
contribution to the transmission decays faster with barrier
thickness compared to non-resonant. We find that for MgO thickness
smaller than 6 MLs the contribution from minority-spin electrons
in the parallel configuration becomes higher than that from
majority-spin electrons. We note that in the calculation by Butler
\emph{et al.} \cite{Butler} this crossover does not occur down to
4 MLs of MgO, although the similar tendency is clearly seen from
Fig.~16 in that paper. This disagreement likely results from the
interface band crossing the Fermi level at a larger distance from
the $\bar\Gamma$ point compared to our calculation.

An important property of the minority-spin interface resonances is
that they strongly contribute to the conductance in the parallel
configuration \emph{only for ideal, symmetric junctions, and only
at zero bias.} Indeed, it is seen in Fig.~\ref{DOSmin} that the
interface DOS for these resonances exceeds the DOS for neighboring
regions of the surface Brillouin zone by one to two orders of
magnitude. Therefore, the interface resonances generate large
tunneling current only if they match similar resonances at the
other side of the barrier. As follows from Fig.~\ref{DOSmin}, a
bias voltage of the order of 0.01~eV is sufficient to destroy this
matching even for ideal epitaxy. We checked this by calculating
the conductance for a small bias voltage using the surface
transmission function (STF) method introduced in
Ref.~\onlinecite{covac}. As expected, at 0.02~eV bias voltage the
conductance becomes fully dominated by majority-spin electrons.
Disorder would also tend to break the matching of the interface
resonances even at zero bias. Therefore, we argue that in real
Fe/MgO/Fe MTJs the minority-spin channel in the parallel
configuration is closed.

Unlike the parallel configuration, the interface resonances
\emph{do} contribute to the conductance in the antiparallel
configuration, where they tunnel into majority-spin states of the
other electrode. The latter have no fine structure in the
Brillouin zone, and hence the conductance is weakly sensitive to a
potential mismatch at the two interfaces which might occur in real
junctions. The enhanced contribution of these interface
resonances, which is clearly seen in Fig.~\ref{transm}c, leads to
the decrease of TMR at low barrier thickness. We emphasize the
fact that although the exact location of the interface resonances
is not determined accurately due to intrinsic limitations of the
density functional theory, their presence at the Fermi level
\cite{Tiusan} inevitably results in the reduced TMR at small
barrier thickness \cite{Yuasa}.

\begin{figure}
\epsfig{file=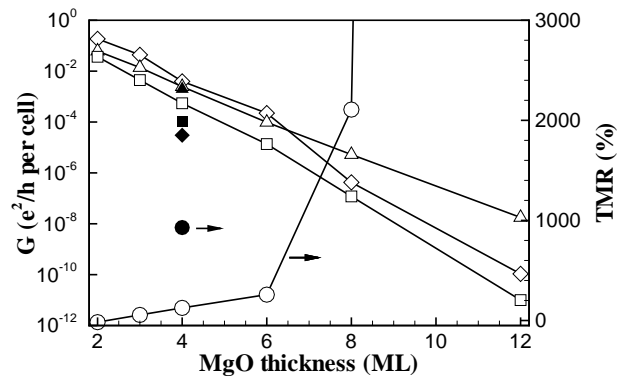,width=0.45\textwidth,clip}
\caption{Conductance (left axis) and TMR (right axis) vs barrier
thickness for Fe/MgO/Fe junctions (open symbols). Triangles:
majority spin, parallel configuration; diamonds: minority spin,
parallel configuration; squares: each spin, antiparallel
configuration; circles: TMR ratio, calculated disregarding
minority spin in the parallel configuration (see text). Solid
symbols: conductance and TMR for a Fe/Ag/MgO/Ag/Fe junction.}
\label{TMR}
\end{figure}

These features are evident in Fig.~\ref{TMR} which shows the
conductance and TMR as a function of barrier thickness. In the
parallel configuration the majority-spin conductance is
controlled by the $\Delta_1$ band which dominates at large
barrier thickness making TMR very large \cite{Butler}. Below 6
MLs of MgO, however, minority-spin electrons overcome the
contribution from majority-spin electrons due to the
contribution from the interface resonances. In the antiparallel
configuration the spin conductance decreases faster than the
majority-spin conductance in the parallel configuration, because
it is dominated by the same interface resonances located away
from the $\bar\Gamma$ point (see Fig.~\ref{DOSmin}). As was
justified above, for real MTJs the minority-spin conductance in
the parallel configuration can be disregarded in the calculation
of TMR. This leads to the increase of TMR with increasing the
barrier thickness. A similar behavior is observed experimentally
until the barrier thickness exceeds approximately 1.5~nm
\cite{Yuasa} which corresponds to 7-8 MLs of MgO. At larger
thickness the rate of decay for the parallel and antiparallel
conductance becomes essentially identical. This crossover may be
due to the loss of $\mathbf{k}_\parallel$ conservation induced
by subbarrier scattering on defects, which makes tunneling
electrons to diffuse over the surface Brillouin zone. The
epitaxial junction model is inapplicable in this regime.

In order to enhance TMR for thin MgO barriers we propose to use
thin epitaxial Ag interlayers deposited at the Fe/MgO interfaces.
Since the lattice parameter of Ag is close to both Fe and MgO
lattice parameters, Ag can be deposited epitaxially on Fe(001)
\cite{Vescovo}, Fe can be grown on Ag \cite{Li}, and Ag on MgO
\cite{Fuchs}. Therefore, epitaxial Fe/Ag/MgO/Ag/Fe(001) tunnel
junctions are feasible. It is known that an epitaxial Ag overlayer
on Fe(001) surface notably modifies the electronic structure of
the surface states \cite{Vescovo}, and it is natural to expect
similar changes for the Fe/MgO interfaces where Fe and MgO
interact only weakly. If the minority-spin interface DOS is
reduced by Ag, the antiparallel conductance will be suppressed. On
the other hand, the majority-spin conductance should not strongly
be affected due to almost perfect transmission through the
Fe/Ag(001) interface \cite{Stiles}. This is the rationale for
using Ag interlayers.

We place 1 ML of Ag atoms on each Fe(001) electrode in the
4-fold hollow sites. The 4 ML MgO barrier is inserted between
Ag-terminated electrodes so that O atoms at the interfacial ML
of MgO lie above the Ag atoms. This interface structure is
considered the most stable for Fe/Ag(001) and Ag/MgO(001)
interfaces \cite{Vescovo,LiWu-Goniak}. To find the equilibrium
interlayer distances, we relax the atomic structure of the MTJ
using the pseudopotential plane-wave method \cite{Payne}
implemented within the Vienna Ab Initio Simulation Package
(VASP) \cite{VASP}. The generalized gradient approximation
\cite{PW} is used for the exchange-correlation energy. We find a
5.2\% reduction in the Fe interlayer distance at the interface,
the distance between the interface Fe and Ag layers being 1.88
\AA, and the distance between Ag and MgO layers being 2.76 \AA.

Figs.~\ref{transm}d-f show the $\mathbf{k}_\parallel$- and
spin-resolved conductance of Fe/MgO/Fe junctions with Ag
interlayers. Not unexpectedly, the majority-spin conductance is
weakly affected by the Ag interlayers, whereas the minority-spin
conductance and the spin conductance in the antiparallel
configuration change dramatically. The most pronounced
difference for the latter two is the disappearance of the
interface resonances that dominated the conductance of the
Fe/MgO/Fe junction with no Ag interlayers (compare
Figs.~\ref{transm}b,c and Figs.~\ref{transm}d,f). This strong
change occurs due to the Fe-Ag hybridization which makes the
interface resonant band more dispersive and hence removes the
Fermi level crossing responsible for the highly conductive
resonant states. A careful examination of the band structure
shows that the interface resonant band still crosses the Fermi
level very close to the $\bar\Gamma$ point (an obscure circular
feature in Figs.~\ref{transm}e,f), but due to its dispersive
nature the interface DOS is small. As a result, this band
crossing contributes 30\% of the total minority-spin conductance
in the parallel configuration, and only about 7\% of the
conductance in the antiparallel configuration. The significant
reduction of the conductance in the antiparallel configuration
leads to dramatic enhancement of TMR which changes from about
130\% to 930\% (see Fig.~\ref{TMR}). Thus, Ag interlayers
practically eliminate the contribution from the interface
resonances and therefore enhance TMR for thin barriers.

In conclusion, we have found that interface resonant states in
Fe/MgO/Fe(001) tunnel junctions contribute to the conductance in
the antiparallel configuration and are responsible for the
decrease of TMR at small barrier thickness, which explains the
experimental results of Yuasa \emph{et al.} \cite{Yuasa}.
Depositing thin Ag interlayers at the Fe/MgO interfaces is an
efficient and practical way to suppress the tunneling
conductance through these resonant states and thereby to enhance
the TMR for thin barriers.

We are grateful to M. van Schilfgaarde and D. A. Stewart for the
use of their computer code. We also thank W. H. Butler and S.
Bl\"ugel for helpful discussions. This work was supported by NSF
(DMR-0203359 and MRSEC DMR-0213808) and Nebraska Research
Initiative. The calculations were performed using the Research
Computing Facility of the University of Nebraska-Lincoln.


\begin{thebibliography}{99}
\bibitem{atomic spheres} For all Fe atoms we use the sphere radius of 2.667 a.u., which
provides space filling in bulk Fe. Inside MgO, we insert an
empty sphere (ES) in the center of each cubic pore formed by 4
Mg and 4 O atoms. The radii are taken to be 2.202 a.u. for Mg,
1.811 a.u. for O, and 1.721 a.u. for ES, providing space filling
in bulk MgO. In addition, ESs are placed at the Fe/MgO interface
exactly above the ESs inside MgO; their position is adjusted to
minimize overlaps. The radii of these interfacial ESs are set to
be 1.761 a.u., as required for global space filling.
\bibitem{CCOR-note} All atomic potential parameters were
obtained using a supercell TB-LMTO calculation with 12 layers of
Fe. To achieve better agreement with the Green's function
calculations, the combined correction term was not used in
TB-LMTO calculations. Our choice of atomic spheres gives
excellent agreement with FP-LMTO calculation for the band gap
(5.86~eV vs 5.82~eV for bulk MgO with the lattice parameter
reduced to match the Fe(001) surface epitaxially).
\end{thebibliography}
\end{document}